\documentclass[twocolumn,aps,showpacs,superscriptaddress,longbibliography]{revtex4-1}

\usepackage{bm}
\usepackage{amsmath}
\usepackage{amssymb}
\usepackage[usenames,dvipsnames]{color}
\usepackage{graphicx}
\usepackage{dcolumn}
\usepackage{hyperref}
\usepackage{natbib}
\usepackage[caption=false]{subfig}
\usepackage{float}
\usepackage{times}
\usepackage{pdfpages}
\usepackage{todonotes}
\hypersetup{
  colorlinks=true,
  citecolor=blue,
  linkcolor=blue,
  urlcolor=blue}

\begin{document}

\title{Electron correlation effects in enhanced-ionization of molecules:
       A time-dependent generalized-active-space configuration-interaction study}
\author{S.~Chattopadhyay}  
\affiliation{Department of Physics and Astronomy,
             Aarhus University, DK-8000 Aarhus C, Denmark}
\author{S.~Bauch}
\affiliation{Institut f\"ur Theoretische Physik und Astrophysik, 
             Christian-Albrechts-Universit\"at zu Kiel, D-24098 Kiel, Germany}
\author{L.~B.~Madsen}
\affiliation{Department of Physics and Astronomy,
             Aarhus University, DK-8000 Aarhus C, Denmark}

\date{\today}

\begin{abstract}
We numerically study models of $\mathrm{H}_2$ and $\mathrm{LiH}$ molecules, aligned collinearly with 
the linear polarization of the external field, to elucidate the possible role of correlation in the 
enhanced-ionization phenomena. Correlation is considered at different levels of approximation with 
the time-dependent generalized-active-space configuration-interaction method. The results of our 
studies show that enhanced ionization occurs in multielectron molecules, that correlation is important 
and they also demonstrate significant deviations between the results of the single-active-electron 
approximation and more accurate configuration-interaction methods. With the inclusion of correlation
we show strong carrier-envelope-phase effects in the enhanced ionization of the asymmetric heteronuclear 
$\mathrm{LiH}$-like molecule. The correlated calculation shows an intriguing feature of
cross-over in enhanced ionization with two carrier-envelope-phases at critical inter-nuclear 
separation. 
\end{abstract}


\maketitle

\section{Introduction}
When a diatomic molecule is exposed to ultra-short laser pulses, the ionization yield is enhanced at 
certain critical inter-nuclear distances. This phenomenon is known as enhanced ionization (EI)  
and has been observed both experimentally~\cite{Constant-PRL-96,Pavicic-PRL-05,Itzhak-PRA-08,
Bocharova-PRL-11,Wu-NComms-12,Lai-PRAR-14,Xu-Arxiv-15} and discussed theoretically
~\cite{Seideman-PRL-95,Zuo-PRAR-95,Chelkowski-JPB-95,Mulyukov-PRA-96,Plummer-JPB-96,Villeneuve-PRA-96} 
(see Ref.~\cite{Bandrauk-12} for a review on EI in molecules). The main interesting feature 
that distinguishes ionization of a diatomic molecule from ionization of an atom is the double-well 
structure of the potential. At the equilibrium inter-nuclear separation, the molecular potential is 
dominated by the Coulombic monopole and in this sense it is similar to the atomic potential. As the 
inter-nuclear separation ${R}$ increases, the internal barrier between the two wells broadens. In 
the presence of a strong electric field the diatomic potential is distorted and the potential changes 
drastically. Since the barrier height in one of the well reduces, the electron can directly tunnel 
to the continuum and this effect enhances the ionization probability at certain critical 
inter-nuclear distance. For one-electron systems, the localization of the wave function in the upper
well of the effective potential formed by the molecular potential and the external field
is a key mechanism as explained in Ref.~\cite{Seideman-PRL-95}. It was shown 
earlier that electron localization in the presence of a strong field is an often encountered 
phenomenon~\cite{Grossman-PRL-91,Bavli-PRL-92}. The probability of localization in one of the  wells
 depends on the laser intensity and frequency as well as on the phase~\cite{Bavli-PRL-92}. 
 It is also important to emphasize that the localization of the electronic 
wave packet depends on the preparation of the initial state. In Ref.~\cite{Zuo-PRAR-95} a different 
mechanism for EI was proposed. In that work EI is seen as the results of the creation of a pair of 
charge-resonant (CR) states which strongly couples to the external field at the critical inter-nuclear 
distance. Along with the creation of CR states the potential experienced by the electron is also 
changed in the presence of the strong field. This is also known as 
charge-resonance-enhanced-ionization (CREI) of diatomic molecules. A one-dimensional model of 
$\mathrm{H}_2$ was studied in Ref.~\cite{Yu-PRA-96} and a full three-dimensional calculation of 
$\mathrm{H}_2$ was presented recently~\cite{Dehghanian-PRAR-10}. 
For a one-electron system like $\mathrm{H}_2^+$, it has been widely discussed from which well of 
the diatomic potential the electron will escape. In Ref.~\cite{Mulyukov-PRA-96}, it is argued that the 
ionization can occur by resonant transition to an excited state supported by the lower well. 
Using $\mathrm{H}_2$, it is also argued that the EI in diatomic molecules 
occurs due to level crossings between the ground state and that of the excited states which 
dissociate into ionic fragments~\cite{Saenz-PRAR-00, Saenz-PRA-02}. The EI in more complex, e.g.,
asymmetric molecules was studied in Ref.~\cite{Kamta-PRL-05,Kamta-PRA-07,Dehghanian-JCP-13}.  
Experimentally the EI was found in molecular iodine~\cite{Constant-PRL-96}, and in molecular 
ions~\cite{Pavicic-PRL-05, Itzhak-PRA-08}. The EI in ${\mathrm{CO}}_{2}$ was experimentally studied 
in Ref.~\cite{Bocharova-PRL-11}, and in Ref. ~\cite{Wu-NComms-12} the tunneling site of the electron 
that takes part in the EI of molecules was experimentally probed. Recent experiments directly 
detected EI in $\mathrm{CO}$ and $\mathrm{N}_{2}$~\cite{Lai-PRAR-14} and 
demonstrated a two peak structure of EI in $\mathrm{H}_2^+$~\cite{Xu-Arxiv-15}.  
 
Solving the time-dependent Schr{\"o}dinger equation (TDSE) directly 
in the presence of a strong external field is possible only for atoms and 
molecules with one or two electrons. Most of the previous calculations therefore use the 
single-active-electron (SAE) approximation. They may, therefore, neglect some important features which 
may arise due to electron correlation. The purpose of this work is to elucidate  EI 
further by investigating in a systematic manner the role of electron correlation on this mechanism.
Several time-dependent many-body methods have been developed to take into account electron 
correlation as accurately as possible. The time-dependent $R$-matrix approach 
~\cite{VanderHart-PRA-07,Lysaght-PRL-08,Lysaght-PRA-09,VanderHart-PRA-14}, the time-dependent
configuration-interaction singles (TDCIS) approximation~\cite{Krause-JCP-05,Rohringer-PRA-06,
Greenman-PRA-10,Karamatskou-PRA-14}, the multiconfigurational time-dependent Hartree-Fock 
method~\cite{Kato-CPL-04,Nest-JCP-05,Caillat-PRA-05,Hochstuhl-JPCS-10,Haxton-PRA-11,Haxton-PRA-12,
Hochstuhl-EPJST-14} and its generalizations~\cite{Miyagi-PRA-13,Sato-PRA-13,Miyagi-JCP-14,
Miyagi-PRA-14,Sato-PRA-15} have been applied to processes involving laser-matter interactions. 
The time-dependent restricted-active-space configuration-interaction (TD-RASCI) 
method~\cite{Hochstuhl-PRA-12}, and time-dependent generalized-active-space (TD-GASCI) 
configuration-interaction method~\cite{Bauch-PRA-14,Larsson-15} systematically take into account the 
electron correlation through a configuration-interaction expansion within a mixed-basis set 
approach with localized and grid-based orbitals to allow for an accurate description of both, 
the bound and continuum states. This method includes the limits of full CI, the 
SAE and the CIS approximations. 
The TD-GASCI method has been used to calculate the photoelectron spectra of one-dimensional 
two and four electron atoms and molecules~\cite{Bauch-PRA-14} and the TD-RASCI method to  
calculate photoionization cross sections of beryllium and neon~\cite{Hochstuhl-PRA-12}. 
In the present work, we use the TD-GASCI method to study correlation phenomena in EI of 
model molecules. First, we consider a model of ${\mathrm{H}}_{2}$. In this homonuclear 
system we study the convergence of the TD-GASCI method by comparing it with fully correlated (``exact'') TDSE 
calculations. To study the correlation trend we use linearly 
polarized laser fields polarized along the inter-nuclear axis within the fixed-nuclei approximation. 
Further we consider a model of ${\mathrm{LiH}}$ and report on EI phenomena within
four electron molecules. In this asymmetric system we study the correlation effects 
in EI and in addition the sensitivity to the relative orientation between the molecule and the
dominant half-cycle of the short pulse as controlled by the carrier-envelope-phase (CEP). 

The paper is organized as follows. In Sec.~\ref{theory}, we briefly present the TD-GASCI method and 
the special basis set used. In Sec.~\ref{numerical_ex}, we use the TD-GASCI method to calculate the 
ionization yield as a function of inter-nuclear distance to study the EI phenomena. We demonstrate
the accuracy of the TD-GAS-CI results of the $\mathrm{H}_2$ model by comparison with TDSE results.
Further we consider calculations for the $\mathrm{LiH}$ model with different GAS partitions corresponding to accounting 
for electron correlation at different levels of approximation. In Sec.~\ref{conc}, we summarize our 
work and conclude.


\section{Theory}
\label{theory}
\begin{figure}
 \includegraphics[width=\columnwidth]{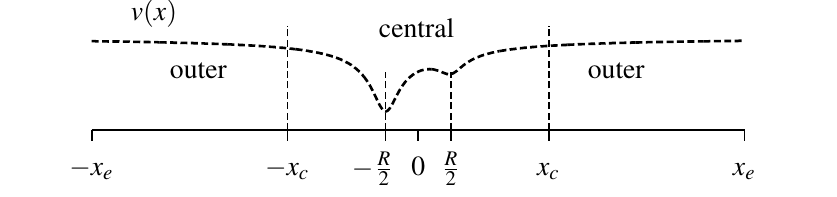}
 \caption{Schematic of the partition-in-space concept. The molecule with binding length $R$ is 
          centered on the grid, the binding potential of a LiH-like model is indicated by $v(x)$ (dashed line). The 
          basis set is partitioned into a central part $|x|<x_c$ and an outer part. For the former, 
          localized Hartree-Fock and pseudo-orbitals for the description of the bound electrons 
          are constructed. Details are given in Ref.~\cite{Bauch-PRA-14}.}
 \label{fig:basis-sketch}
\end{figure}

Here we briefly recall the TD-GASCI methodology~\cite{Bauch-PRA-14}. To ease notation we consider 
a diatomic molecule with nuclei at $\pm{R/2}$ and charges $Z_1$ and $Z_2$, respectively. The TDSE 
for $N_{el}$ -electrons with fixed nuclei then reads (we use atomic units throughout this work)
\begin{equation}
      i\frac{\partial}{\partial t}|\Psi(t)\rangle = H(t)|\Psi(t)\rangle \; .
\end{equation}
The time-dependent Hamiltonian is given by 
\begin{equation}
    H(t)=\sum_{i=1}^{N_{el}} {h}_{i}(t)+\sum_{i < j}^{N_{el}} w_{ij} \; ,
    \label{td_ham}
\end{equation}
with the one-body part of electron $i$ given by
\begin{equation}
 {h}_i(t)=-\frac{1}{2}\nabla_i^2 - \frac{Z_1}{|\bm{r}_i - \frac{\bm{R}}{2}|} - 
           \frac{Z_2}{|\bm{r}_i + \frac{\bm{R}}{2}|} +  \bm{r}_i\cdot \bm{F}(t) \;,
           \label{eq:h1body}
\end{equation}
with $\bm{F}(t)$ is the laser field. In Eq. \eqref{td_ham} the two-body interaction is given by
\begin{equation}
 w_{ij} = \frac{1}{|\bm{r}_i - \bm{r}_j|} \;.
\end{equation}
The many-body wave function is expanded into a basis of Slater determinants $|\Phi_I\rangle$,
\begin{equation}
     |\Psi(t)\rangle=\sum_{I \in {\cal V}_{\text{Exc}} }C_I(t) | \Phi_I\rangle \;,
     \label{full_ci_exp}
\end{equation}
with time-dependent expansion coefficients $C_I(t)=\langle \Phi_I|\Psi(t)\rangle$ and $I$ is a
multi-index specifying the configurations drawn from the Hilbert space ${\cal V}_{\text{Exc}}$.  
The Slater determinants themselves are constructed from $N_b$ time-independent single-particle 
orbitals $\varphi_i$ ($2N_b$ spin orbitals), within a mixed-basis set approach. Thereby, the space 
variable $\bm{r}$ is partitioned into a central region $|\bm{r}|\leq |\bm{r}_c|$, where localized 
Hartree-Fock (HF) and pseudo-orbitals are used to describe the bound electrons, and an outer region 
$|\bm{r}_c|<|\bm{r}|\leq |\bm{r}_e|$ with a finite-element discrete-variable-representation (FE-DVR) 
basis description for the outgoing electron wave, see Fig.~\ref{fig:basis-sketch} for a schematic 
of the 1D situation ($x$ corresponds to one component of $\bm{r}$).
Details of the basis and its implementation are given in 
Ref.~\cite{Bauch-PRA-14}.

The corresponding matrix form of the TDSE then reads as
\begin{equation}
     i\frac{\partial}{\partial t} C_I(t) = \sum_{J \in {\cal V}_\text{Exc}}   H_{IJ}(t) C_J(t) \;,
    \label{fullci_tdse}
\end{equation}
with the Hamiltonian matrix $H_{IJ}(t)=\langle \Phi_I |H(t) | \Phi_J \rangle$. This matrix can be 
constructed by evaluating and partially transforming the associated one- and two-electron integrals. 
If the sums in Eqs.~\eqref{full_ci_exp} and \eqref{fullci_tdse} run over all possible excitations 
${\cal V}_{\text{Exc}}$, the full CI (FCI) method~\cite{Szabo-MQC-96} is obtained and an exact 
solution of the TDSE within the limits of the finite single-particle basis set is retrieved.
Typically, an FCI expansion of the TDSE is numerically intractable for processes involving one or 
more continua as in photoionization, and for systems consisting of more than two active electrons 
due to an exponential scaling in the number of configurations with the number of basis functions. 
The generalized-active-space (GAS) concept aims to overcome this exponential barrier by choosing 
the most relevant Slater determinants for the dynamics under consideration and thus forming a subset 
of the FCI many-particle basis set, ${\cal V}_\text{Exc}={\cal V}_\text{GAS}$ in 
Eq.~\eqref{full_ci_exp}. By systematically including 
more Slater determinants, convergence of the method towards the fully correlated result is obtained.

\begin{figure}
  \includegraphics[width=\columnwidth]{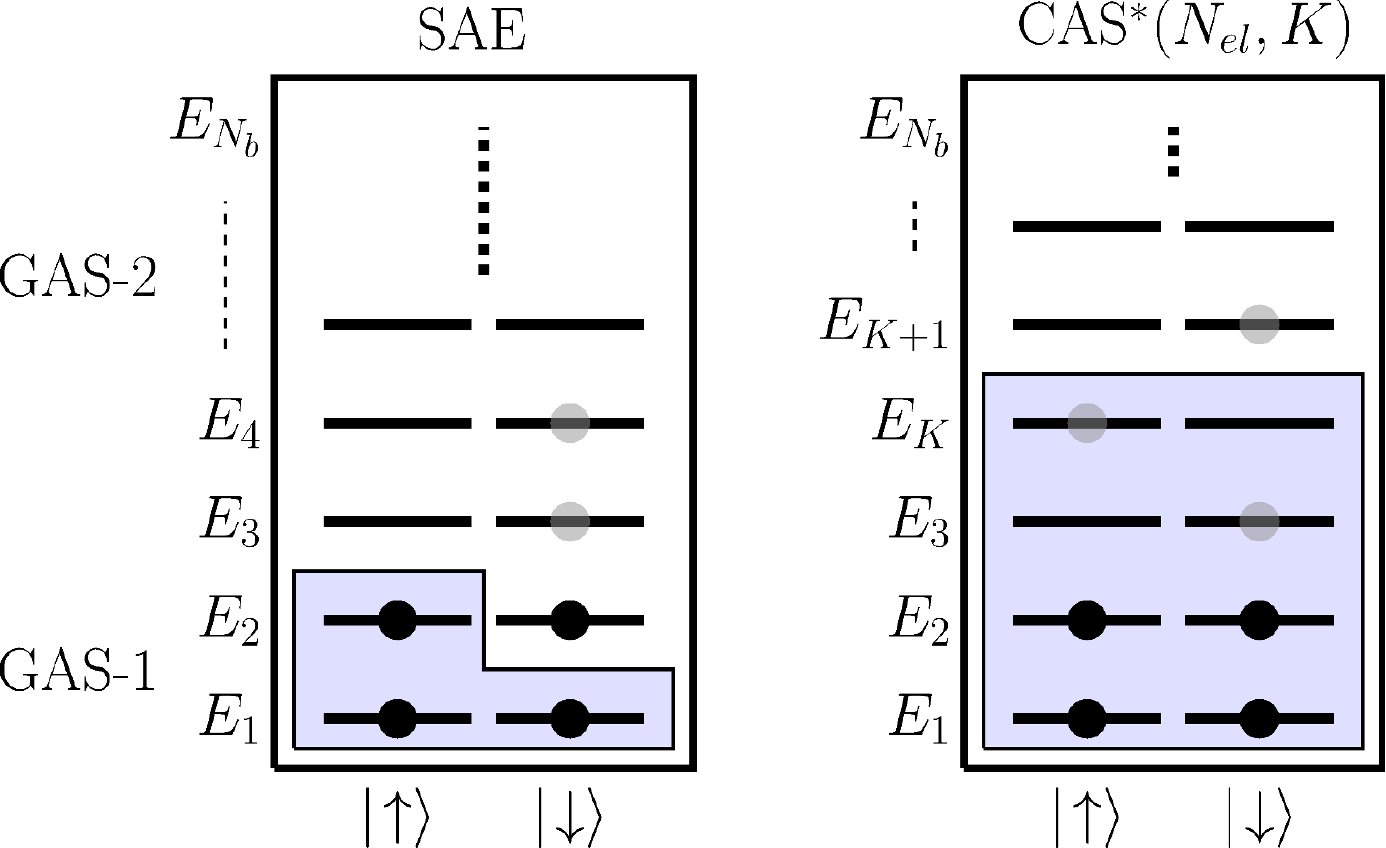}
  \caption{(color online). Generalized active spaces used in this work. The energies of the 
           single-particle levels (up to $N_b$ spatial orbitals) are denoted by $E_i$. The two spin 
           configurations, $|\hspace{-0.1cm}\uparrow\rangle$ and $|\hspace{-0.1cm}\downarrow\rangle$,
           are degenerate in the spin restricted HF ansatz. The left panel shows the SAE approximation, 
           where only one electron ($\downarrow$ in level $E_2$ in this case) is allowed to be excited. 
           The right panel shows the complete-active space (CAS$^*$) situation, where correlation 
           can be systematically built in with the incorporation of more and more active orbitals. 
           $N_{el}$-electrons are allowed to occupy $K$ spatial orbitals and one electron excitations 
           for any spin configuration are allowed above level $E_K$.}
 \label{fig:gasscheme}
\end{figure}
The relevant GAS partitions for this work are summarized in Fig.~\ref{fig:gasscheme}. The simplest 
approximation is the SAE approximation, in which only one selected electron may occupy excited 
Slater determinants, likewise the CIS approximation can be obtained from the GAS. By construction, 
no additional potentials are needed to be evaluated to obtain this widely used truncation. The other 
partition refers to the complete-active-space (CAS) concept~\cite{Olsen-JCP-88,Helgaker-MEST-14}, 
which corresponds to a FCI description (with possibly a frozen core) up to a spatial orbital index 
$K$. In the present TD-GASCI approach, the CAS is accompanied by additional single excitations out 
of the CAS, indicated in our notation by CAS$^*(N_{el},K)$, where $N_{el}$ denotes the active 
electrons and $K$ the number of spatial orbitals within the CAS.  In this context we wish 
to emphasize that we consider only one electron in the continuum throughout and double ionization 
is neglected, per construction of the GAS~\cite{Bauch-PRA-14}.

\section{Correlation effects in enhanced ionization}
\label{numerical_ex}
To study EI in molecules, we consider colinear models of $\mathrm{H}_2$ and $\mathrm{LiH}$.
We use the regularized Coulomb potential  between the electrons and the nuclei,
\begin{equation}
     v(x,R)=-\frac{Z_1}{\sqrt{(x-\frac{R}{2})^2+s^2}}-\frac{Z_2}{\sqrt{(x+\frac{R}{2})^2+s^2}} \;.
\end{equation}
Here  $x$ is the electron coordinate, $R$ is the inter-nuclear distance, and $Z_i$ $(i=1,2)$ the 
nuclear charges. For all calculations we use the softening parameter, $s=1$.  A two-electron
model of $\mathrm{H}_2$ is defined by $Z_1=Z_2=1$ and a four electron model of $\mathrm{LiH}$ 
by $Z_1=1$, and $Z_2=3$. The electron-electron interaction in this model reads
\begin{equation}
     w(x_i,x_j)=\frac{1}{\sqrt{(x_i-x_j)^2+s^2}} \;.
\end{equation}
\begin{figure}
 \includegraphics[width=\columnwidth]{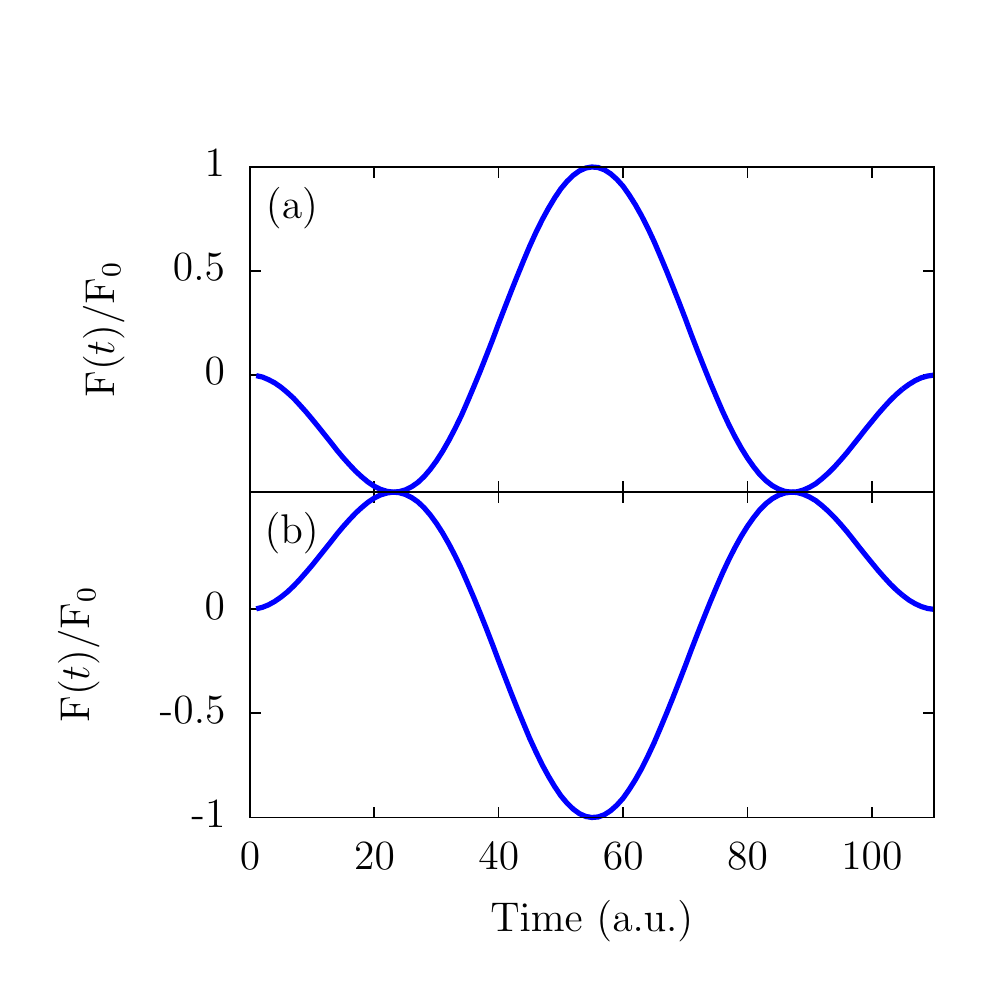}
 \caption{(color online). Normalized electric field with  (a) $\varphi_{\textup{CEP}} = 0 $ 
           and (b) $\varphi_{\textup{CEP}} = \pi$.}
 \label{fig:E_field}
\end{figure}
 Such kind of models for diatomic molecules are well established in strong-field physics involving 
near-infrared fields~\cite{Kulander-PRA-96}.
In the calculations we use the length gauge to describe the interaction with the laser field. We 
use the vector potential with a sine-square envelope~\cite{Han-PRA-10},
\begin{equation}
   A(t)  = \frac{F_0}{\omega}\sin^2\Big(\frac{\pi t}{T}\Big) \sin{(\omega t + \varphi_{\textup{CEP}})} 
            \qquad (0 \le t \le T) \;,
   \label{vector_pot}
\end{equation}
where $T = n \frac{2\pi}{\omega}$ is the pulse duration with $n = 1$, the number of cycles.
The corresponding electric field is $F(t) = - \frac{\partial A(t)} {\partial t}$. $F_0$ is the 
maximum amplitude of the laser field to be specified below in the calculations, $\omega$ is the 
angular frequency and $\varphi_{\textup{CEP}}$ is the carrier-envelope-phase (CEP). 
The form of the laser pulse is shown in Fig.~\ref{fig:E_field}. For the present work we use 800 nm 
($\omega = 0.057$) and single-cycle pulses of duration $2.65$ fs ($110$ a.u.). The vector 
potential in Eq.~\eqref{vector_pot} ensures that the time-integral over the electric field is
vanishing, as it should be~\cite{Madsen-PRA-02}. We note that the pulse duration considered here
is so short that a molecule will not have time to dissociate to the range of critical 
inter-nuclear distances for EI during the presence of the pulse. Experimentally a 
pump-probe scheme is therefore required to investigate the EI phenomena for such short pulses.

To extract the total ionization probability, we add a complex absorbing potential (CAP) to 
the Hamiltonian,
\begin{equation} 
       H^{\textup{CAP}}(t)=H(t)-iV_{\textup{CAP}} \;,
\end{equation}
where the CAP has the form~\cite{Zanghellini-JPB-06}
\begin{equation}
      V_{\textup{CAP}}(x)=1-\cos\left(\frac{\pi( |x|-x_{\textup{CAP}})}{2(x_s-x_\textup{CAP})}\right) \;,
      \label{cap}
\end{equation}
for $|x|>x_{\textup{CAP}}$ and zero otherwise. For sufficiently long propagation time, $t_f$, after 
the end of pulse, the continuum part of the wave function has passed into the region of space beyond 
$x_\textup{CAP}$ and has been absorbed. The total ionization yield is then given by 
\begin{equation} 
    \mathcal{P}(t_f)=1-\mathcal{N}(t_f) \;, 
    \label{ionization-prob}
\end{equation}
with $\mathcal{N}(t_f) = \langle \Psi(t_f)|\Psi(t_f) \rangle$~\cite{Kulander-PRA-87}. 
To extract the ionization yield after the end of the pulse, we propagate the equations of motion to 
a final time, $t_f$ = 241 fs. We found that this time is sufficient to obtain converged results, 
also for correlated situations.

We partition the full simulation box into an inner (central) and an outer region. In the present 
simulation the main focus is on a wide range of inter-nuclear separations. Therefore, we need a relatively 
large box for the central region, $|x| < x_c$, as the correlated bound part of the wave function is described 
by localized HF and pseudo-orbitals. We use $x_c=16$ in the present 
calculation, cf. Fig.~\ref{fig:basis-sketch}, and checked carefully for convergence of our results.
We use 16 elements which have 8 DVR functions each for the central region which results in 
total 111 FE-DVR functions in the central region for all calculations. 
Further increasing the number of DVR functions leads to almost 
no change in the HF energy. The TD-GASCI calculation is done on the full simulation box to take 
into account the continuum. In the present calculation we set the box size to $x_e=200$, and the 
outer region is described by the FE-DVR functions. By using the same FE-DVR parameters as for
the inner region, we therefore include 699 FE-DVR functions for the whole simulation box. 
We checked the convergence with respect to different box parameters and the above parameters gave 
converged results.

\subsection{Two-electron $\mathrm{H}_2$-like molecule}
\label{ssec:h2model}
\begin{figure}
 \includegraphics[width=\columnwidth]{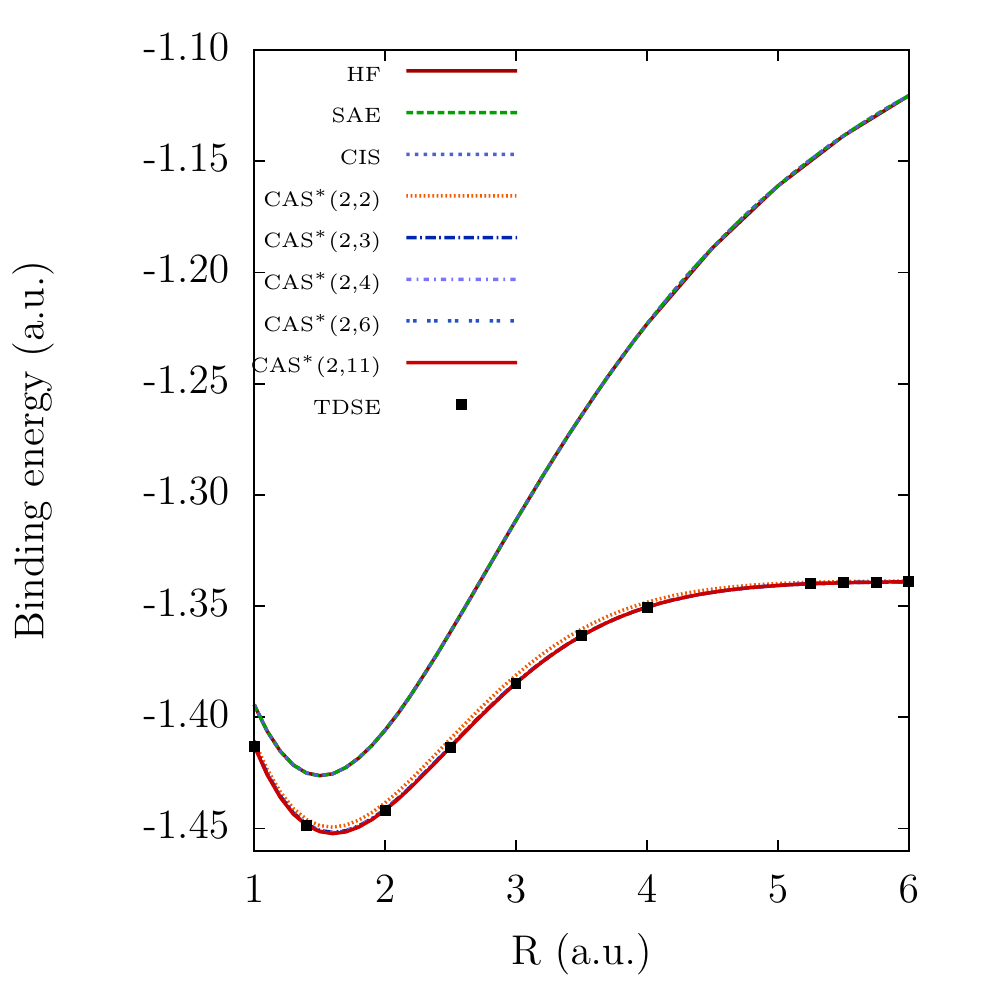}
 \caption{(color online). Total ground-state energy of the 2-electron $\mathrm{H}_2$-like molecule. 
          In the upper curve the SAE and the CIS approximation overlap with the HF result. 
          CAS$^*(2,2)$ represents the TD-GASCI calculation with two active orbitals in the GAS.
          Similarly  CAS$^*(2,11)$ represents the TD-GASCI calculation with eleven active orbitals
          in the GAS.}
 \label{fig:H2-binding}
\end{figure}
The two electron system is the preferred choice of system to use to validate the approach because 
we can directly compare the results of the TD-GASCI simulations with TDSE simulations, which are
performed in the same FE-DVR basis set.
We begin with the ground state energy calculation with imaginary time propagation to demonstrate 
the features of the TD-GASCI method. In Fig.~\ref{fig:H2-binding} we show the total ground state 
energy of 1D  $\mathrm{H}_2$ with HF, SAE, CIS approximations and different GAS methods
along with the TDSE calculations. Let us first discuss the 
approximations including singly excited determinants only. In the SAE approximation, the wave function reads
\begin{equation}
   |\Psi_{\rm SAE}(t)\rangle = C_0(t)|\Phi_0\rangle + \sum_{a}C_{i}^{a}(t)|\Phi_i^a\rangle \;.
\end{equation}
Here, $|\Phi_0\rangle$ is the HF reference determinant, and $|\Phi_i^a\rangle$ is a singly-excited
determinant with the time-dependent CI-coefficients. In this approximation the sum runs over all
virtual orbitals, $a$, with a fixed core orbital, $i$. Therefore, it represents the effective
interaction felt by the electrons from the ionic core. In the CIS approximation the wave function 
reads
\begin{equation}
   |\Psi_{\rm CIS}(t)\rangle = C_0(t)|\Phi_0\rangle + \sum_{i,a}C_{i}^{a}(t)|\Phi_i^a\rangle \;.
\end{equation}
This approximation incorporates all possible single excitations from the core orbitals to the virtual
orbitals. The time-dependent CIS approximation has been widely used for understanding the processes involving
strong-field ionization~\cite{Rohringer-PRA-06,Greenman-PRA-10,Karamatskou-PRA-14}. Although these 
two approximations take into account a certain contribution to the correlation by accounting for the singly-excited 
determinants, the SAE and CIS approximations do not improve the ground state energy. This 
is according to Brillouin's theorem~\cite{Szabo-MQC-96}, which states that the matrix elements of 
the Hamiltonian between the reference state and singly-excited states vanish within a HF basis. 
From Fig.~\ref{fig:H2-binding} we can see that the ground state energy curve obtained from the SAE 
and CIS approximations overlap with the HF ground state.  
Further, these approximations are unable to reproduce the correct dissociation limit due to the restricted
HF ansatz used in this work, which prohibits dissociation into two open shell systems.
For the GAS methods, the wave 
function is written as 
\begin{eqnarray}
   |\Psi_{\rm GAS}(t)\rangle & = & C_0(t)|\Phi_0\rangle + \sum_{ia}C_{i}^{a}(t)|\Phi_i^a\rangle  \nonumber \\
                &  & + \frac{1}{4} \sum_{ijab}C_{ij}^{ab}(t)|\Phi_{ij}^{ab}\rangle  + \cdots  \;.
   \label{gas-wf}
\end{eqnarray}
Let us consider the CAS$^*(2,2)$ model with two active orbitals in the lowest GAS partition. It 
means, the summation over the core orbitals in Eq.~\eqref{gas-wf} runs from one to two. Similarly, 
in the CAS$^*(2,3)$ model we have three active orbitals and so on. With the CAS$^*(2,2)$ model 
Fig.~\ref{fig:H2-binding} shows that the ground state energy curve improves significantly over 
the SAE and CIS approximations and it tends to converge towards the exact TDSE ground state. We 
observe that the CAS$^*(2,4)$ model with four active orbitals in the lowest GAS partition produces 
a converged result. In order to check the convergence, we further increased the number of active 
orbitals to six and then to eleven and found that the  CAS$^*(2,4)$ model is already converged.
This is due to the multi-reference character of the CAS wave functions, which allows for the 
break up into two H atoms with singly-occupied orbitals. This is in contrast to the 
underlying restricted HF method, where orbitals are always doubly occupied.

\begin{figure}
    \includegraphics[width=\columnwidth]{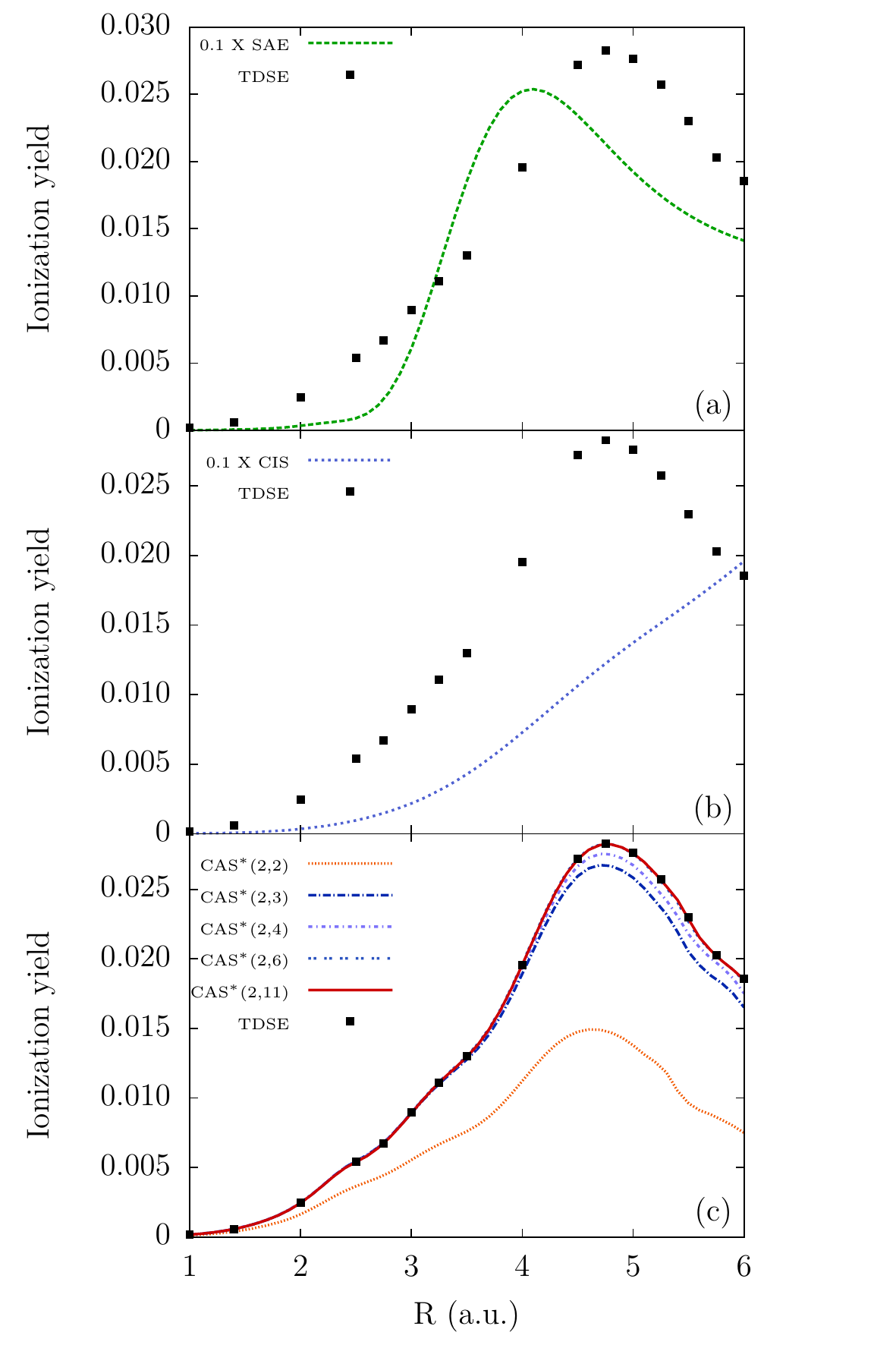}
    \caption{Ionization yield vs inter-nuclear distance $R$ for our model $\mathrm{H}_2$ with 
             ${F}_0 = 0.05$ for (a) SAE approximation, (b) CIS approximation, and  (c) 
             GAS approximations. The CAS$^*(2,2)$  model represents two active orbitals in the lowest GAS partition and similarly,
             e.g, the CAS$^*(2,11)$ model incorporates eleven active orbitals in the lowest 
             GAS partition. The squares are fully correlated TDSE reference data.}
     \label{fig:H2-ionization-05}
\end{figure}
After the  preparation of  the $\mathrm{H}_2$-like molecule in its ground state 
by imaginary time propagation, we use the 800 nm laser pulse of Fig.~\ref{fig:E_field} with 
$\varphi_{\textup{CEP}} = 0$ and two different field strengths, (i) ${F}_0$ = 0.05
(8.75$\times 10^{14}$ W/cm$^2$) and (ii) ${F}_0$ = 0.15 (7.87$\times 10^{15}$ W/cm$^2$), 
to ionize the  molecule. Since the $\mathrm{H}_2$ model is homonuclear, the 
$\varphi_{\textup{CEP}} = \pi$ field of Fig.~\ref{fig:E_field}(b) gives the same results 
for the ionization yield as $\varphi_{\textup{CEP}} = 0$. In Fig.~\ref{fig:H2-ionization-05} we show the 
ionization yield as a function of the inter-nuclear separation. In Fig.~\ref{fig:H2-ionization-05}(a) 
we compare the results between the SAE approximation and TDSE calculations. We have scaled down the 
SAE results by an order of magnitude. With the increase of the inter-nuclear distance, the ionization yield 
enhances in both models and then it gradually decreases. In the SAE approximation the EI peak is 
at $R = 4.1$ which is well below the exact TDSE result of $R = 4.7$. An earlier 
calculation on a similar 1D model observed the EI peak around the same inter-nuclear distance 
~\cite{Yu-PRA-96}. The SAE approximation captures the overall behavior of EI but it overestimates 
largely the magnitude and peaks at the wrong position and, therefore, demonstrates the failure of 
the model to describe the EI phenomena quantitatively. The comparison between the CIS approximation 
and the TDSE calculation is shown Fig.~\ref{fig:H2-ionization-05}(b). Similarly to SAE, the CIS yield
is multiplied by $0.1$. The CIS approximation is unable to capture the feature of EI with the laser 
parameters used in the present calculation. It can describe the features qualitatively at the 
equilibrium separation, but the EI phenomena which occurs at large inter-nuclear separation can 
not be described. 

In order to understand the electron correlation effects we turn to the discussion of 
Fig.~\ref{fig:H2-ionization-05}(c). The lower curve corresponds to the CAS$^*(2,2)$ model, i.e.,  
two active orbitals in the lowest GAS partition, and one can see that this model predicts the 
correct EI behavior but the magnitude is much less than the corresponding TDSE value. Unlike the 
SAE and CIS approximations, the CAS$^*(2,2)$ model nevertheless captures the correct position of 
the EI peak. By successively increasing the number of orbitals [CAS$^*(2,3)$ to CAS$^*(2,11)$ in 
Fig.~\ref{fig:H2-ionization-05}], we achieve convergence of the
ionization yield and observe perfect agreement with the TDSE calculations (squares).
We find that at least six active orbitals in the GAS partition are needed to 
describe the EI phenomena quantitatively. 

\begin{figure}
     \includegraphics[width=\columnwidth]{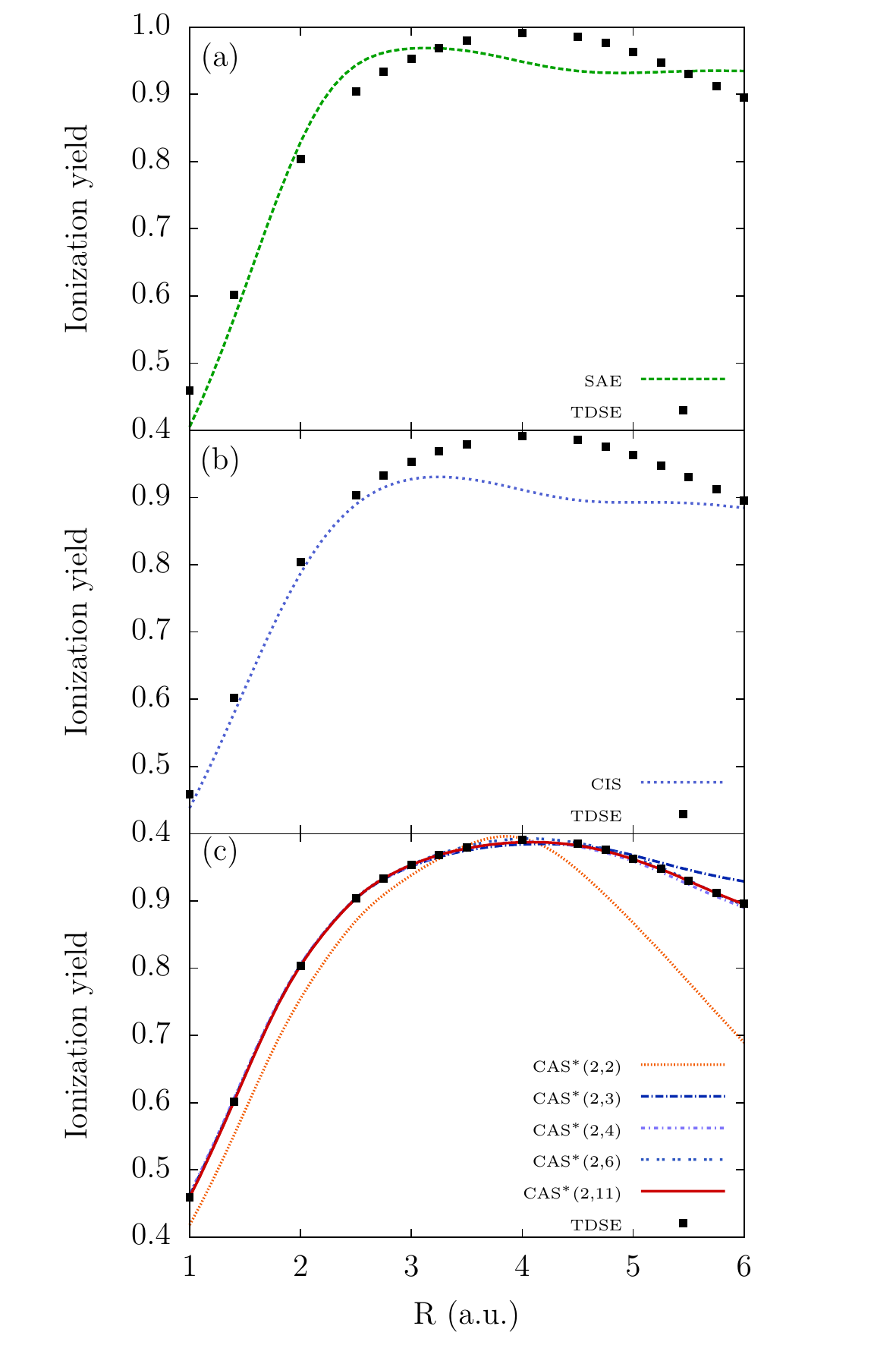}
     \caption{The same as Fig.~\ref{fig:H2-ionization-05}, but for ${F}_0 = 0.15$.}
     \label{fig:H2-ionization-15}
\end{figure}

In the next step, we increase the laser intensity to ${F}_0$ = 0.15 and observe that the 
CAS$^*(2,6)$ model with six active orbitals is again the smallest GAS partition where the results 
converge to the TDSE results, see Fig.~\ref{fig:H2-ionization-15}.  The idea of introducing the 
GAS concept is to mitigate the computational constraint arising due to exponential scaling. By 
comparing with the exact TDSE results, we found, as expected, an enormous gain in terms of 
computational costs. To give an example, the converged CAS$^*(2,6)$ calculations took 35 CPU 
hours on a 2.8 GHz Intel Ivy bridge CPU while the TDSE needed about 1400 CPU hours on an AMD Opteron 
CPU. Albeit the drastic reduction in CPU time, the TD-GASCI method with a limited number of active 
orbitals is still able to describe quantitatively the EI phenomena, including correlation.   

\subsection{Four-electron $\mathrm{LiH}$-like molecule}
\begin{figure}
 \includegraphics[width=\columnwidth]{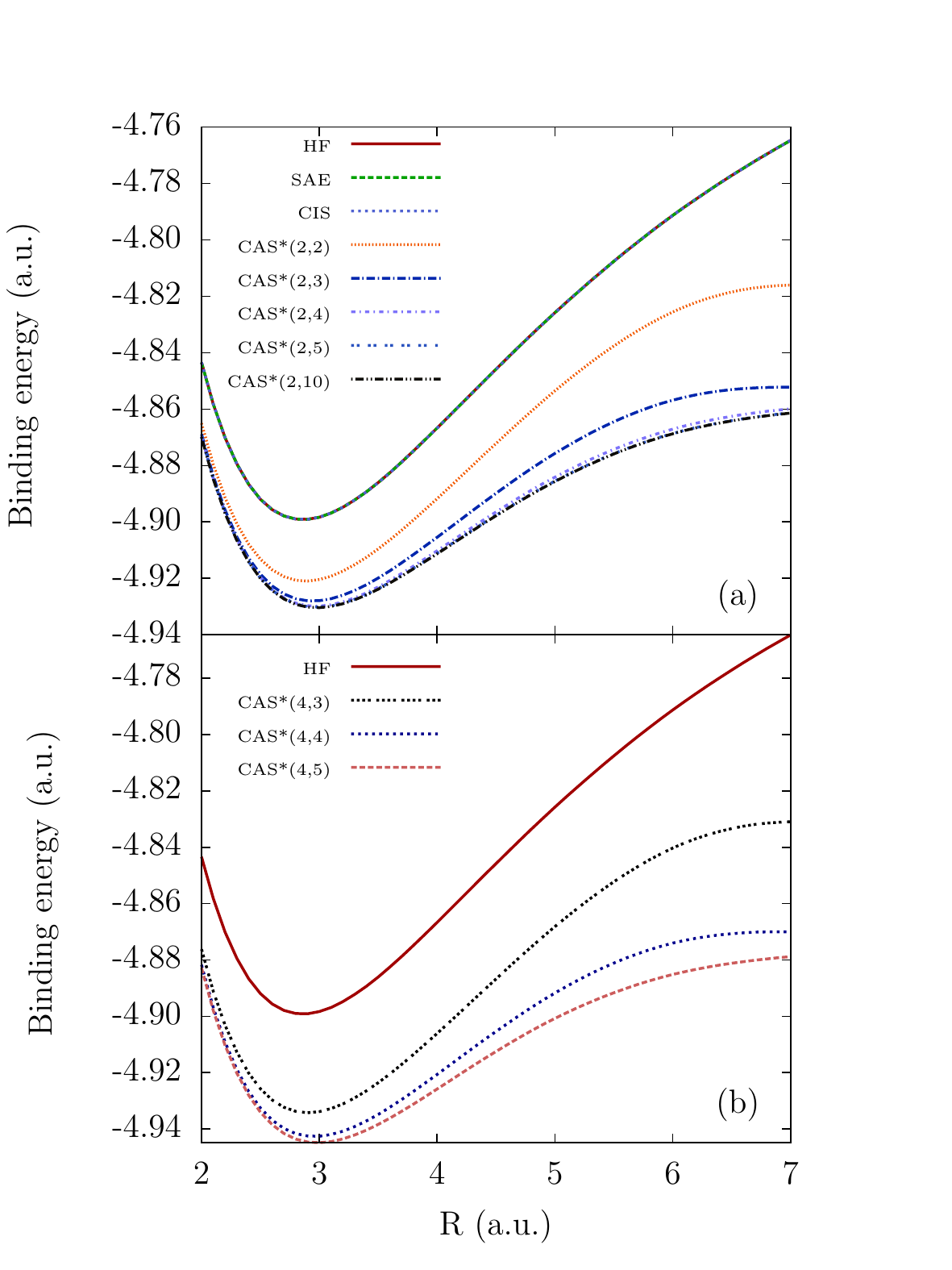}
 \caption{(color online). Binding energy of the $\mathrm{LiH}$ model. (a): HF, SAE and CIS approximations 
          together with GAS calculations [CAS$^*(2,\bullet)$] with two active electrons, (b): the same but with four active
          electrons in the CAS [CAS$^*(4,\bullet)$].}
 \label{fig:LiH-binding}
\end{figure}
We now turn our attention to the four-electron heteronuclear $\mathrm{LiH}$ molecule for which TDSE simulations of
EI are infeasible. $\mathrm{LiH}$ lacks the reflection symmetry of $\mathrm{H}_2$ at the origin and thus strong CEP for few-cycle
pulses can be expected.

We choose the geometric  center as the origin of the 
coordinate system and choose $Z_1=1$ and $Z_2=3$ in Eq.~\eqref{eq:h1body}, i.e.,  the $\mathrm{Li}$ atom is 
at $-R/2$ and the $\mathrm{H}$ atom is placed at $+R/2$.
We prepare the $\mathrm{LiH}$-like molecule in its ground state by 
imaginary time propagation. All numerical parameters are the same as for $\mathrm{H}_2$, see Sec.~\ref{ssec:h2model}.

\subsubsection{Ground-state energies}
The ground state energy curve for $\mathrm{LiH}$ is shown in Fig.~\ref{fig:LiH-binding}. In 
Panel (a) the HF, SAE and CIS approximations with CAS$^*(2,\bullet)$ approximations
involving two active electrons are shown. 
Similar to $\mathrm{H}_2$, the multi-reference description is essential and SAE and CIS
fail to reproduce the correct dissociation limit. 
We find that a CAS$^*(2,5)$ is sufficient for two active electrons,
as seen by comparison with the  CAS$^*(2,10)$ calculation. 
To test the frozen-core approximation, we use 
four active electrons and obtain the converged result for the ground state energy, as shown in 
panel (b). As expected, the energy is lowered but the equilibrium distance is around the same internuclear distance.
Similar to Ref.~\cite{Bauch-PRA-14}, the CAS$^*(4,5)$  with five 
active orbitals produces the converged ground state energy curve.

\begin{figure*}
     \includegraphics[width=14cm]{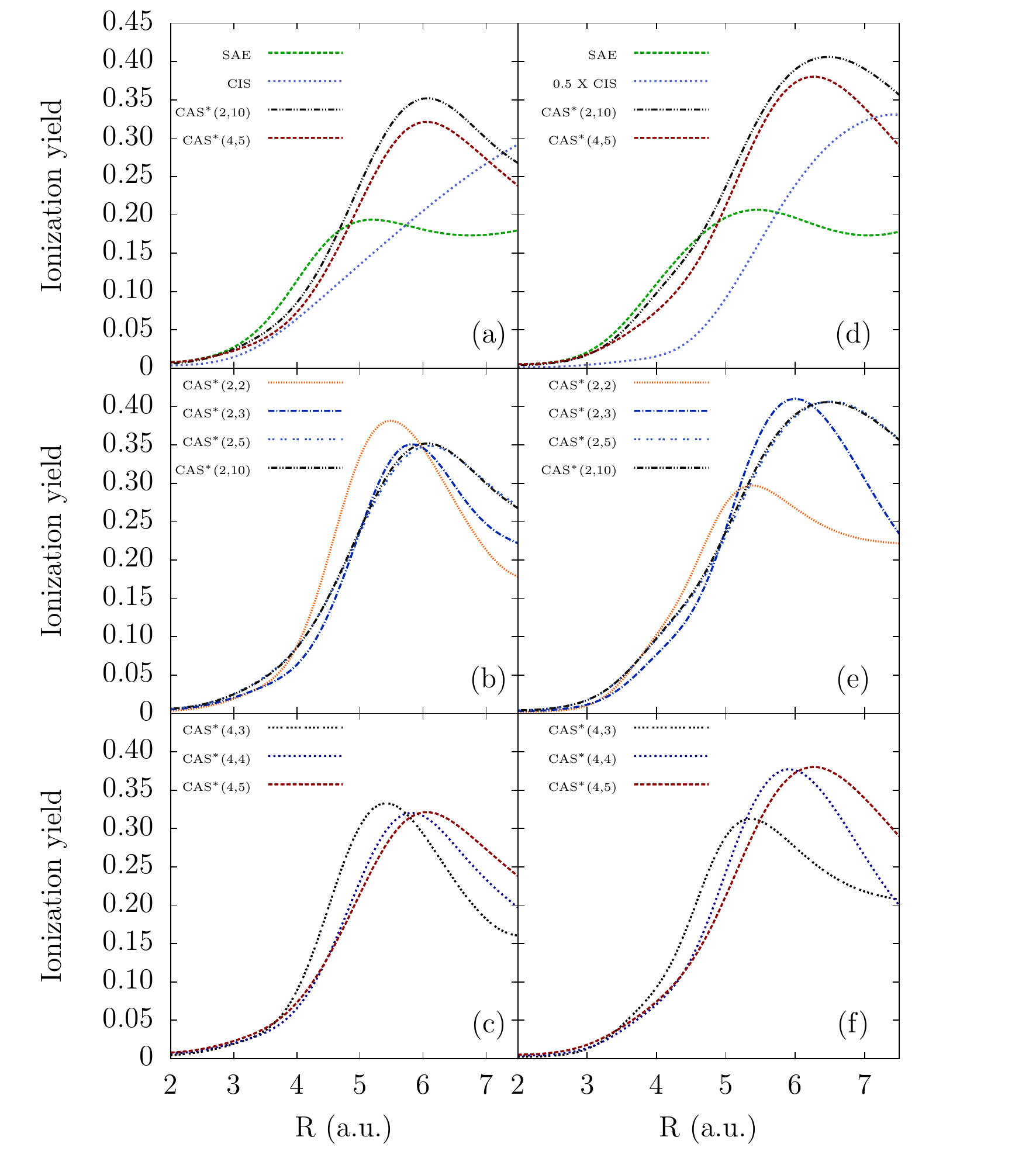}
     \caption{Ionization yield vs inter-nuclear distance $R$ for  $\mathrm{LiH}$ with 
              ${F}_0 = 0.025$. In panels (a-c), $\varphi_{\textup{CEP}} = 0$, 
              for (d-f) $\varphi_{\textup{CEP}} = \pi$ (see Fig.~\ref{fig:E_field}). (a) and (d) Comparison 
              between SAE, CIS approximations and fully converged GAS methods with two and four active 
              electrons. (b) and (e) Comparison between different GAS methods with two active electrons. 
              (c) and (f) Comparison between different GAS methods with four active electrons.}
     \label{fig:lih-ionization-025}
\end{figure*}

\subsubsection{Enhanced ionization}
In order to understand the correlation effects in EI phenomena, we expose the $\mathrm{LiH}$ molecule 
to a laser pulse with field strength ${F}_0 = 0.025$ (2.18$\times 10^{13}$ W/cm$^2$) for
 $\varphi_{\textup{CEP}} = 0$ and $\varphi_{\textup{CEP}} = \pi$ [Fig.~\ref{fig:E_field}]. This 
 choice is equivalent to the two possible orientations in 1D of the molecule
w.r.t. the dominant half-cycle.
In Figs.~\ref{fig:lih-ionization-025}(a-c), we show the EI calculation with $\varphi_{\textup{CEP}} = 0$. 
We compare the results of the SAE, CIS approximations and two 
GAS methods with two and four active electrons in Fig.~\ref{fig:lih-ionization-025}(a). The SAE 
approximation shows the EI peak at $R = 5.2$. Like in the case of $\mathrm{H_2}$, the SAE 
approximation predicts the EI phenomena for $\mathrm{LiH}$, but again the CIS approximation fails 
to describe the EI phenomena. In Fig.~\ref{fig:lih-ionization-025}(b), we show different GAS 
calculations with two active outer electrons. The CAS$^*(2,2)$ model does not predict 
the correct EI peak position and it overestimates the magnitude of the ionization yield. Further 
increasing the number of active orbitals to three we found that the EI peak is shifted towards 
larger inter-nuclear separation. We systematically increased the number of active orbitals and 
found that we need at least five active orbitals to obtain converged results for EI. The converged 
EI peak is at $R = 6.1$ with the CAS$^*(2,5)$ model. To check this 
convergence we increase the number of active orbitals to ten and the observed yield shows a trend 
of convergence. In Fig.~\ref{fig:lih-ionization-025}(c) we show the results with four active 
electrons in the GAS partition and the CAS$^*(4,5)$ model with five active orbitals in the GAS 
partition leads to a converged result. The CAS$^*(4,5)$ model predicts the same qualitative behavior 
like the CAS$^*(2,5)$ model. The EI peak in the  CAS$^*(2,5)$ and the CAS$^*(4,5)$ models have the 
same position but the magnitude is different. In terms of computational cost the CAS$^*(4,5)$ model 
is much more expensive than CAS$^*(2,5)$. To give an example the CAS$^*(4,5)$ model took 530 CPU 
hours compared to 24 hours of the CAS$^*(2,5)$ model on a 2.8 GHz Intel Ivy bridge CPU. Thus using 
only two active electrons in the GAS partition gives us more than a 20 times reduction in the 
computational cost, a tremendous gain. 
 
In Figs.~\ref{fig:lih-ionization-025}(d-f), we show the EI calculation with 
$\varphi_{\textup{CEP}} = \pi$, and the conclusions w.r.t. convergence are the same as for 
Figs.~\ref{fig:lih-ionization-025} (a-c).
As we incorporate more correlation the yield is converged with five active orbitals. In 
Ref.~\cite{Bauch-PRA-14} it was shown that the ionization yield at equilibrium is larger for 
$\varphi_{\textup{CEP}} = 0$ compared to $\varphi_{\textup{CEP}} = \pi$ . We observe that the 
ionization yield with $\varphi_{\textup{CEP}} = \pi$ is consistently greater than with  
$\varphi_{\textup{CEP}} = 0$  at the critical inter-nuclear separation in the different GAS 
calculations. It certainly illustrates the importance of correlation contributions in the direction 
of electron emission from a polar molecule like $\mathrm{LiH}$.

\begin{figure}
 \includegraphics[width=\columnwidth]{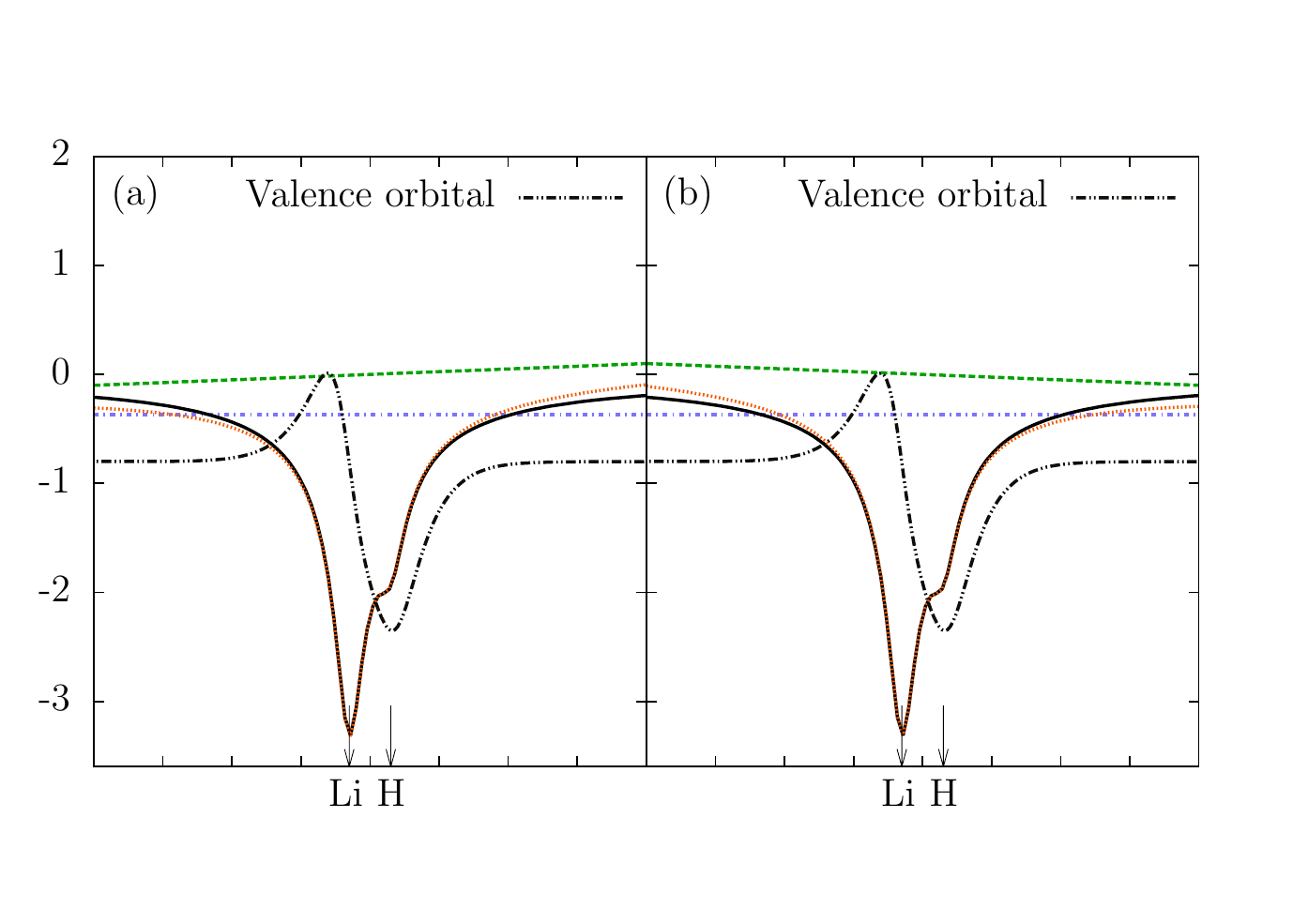}
 \caption{(color online). Sketch of the strong-field ionization scenarios of the $\mathrm{LiH}$ model.  
          Shown are the binding potential (black solid lines) and the modified potential (red dotted lines)
          by the electrical field (green dashed lines) at the peak of the single-cycle pusle F(t) along
          with the valence orbital (black dashed dotted). The orbital energy is indicated by the blue horizontal
          line. Panel (a) is for $\varphi_{\textup{CEP}}$ = 0, (b) for $\varphi_{\textup{CEP}} = \pi$
          (see Fig.~\ref{fig:E_field}).}
 \label{fig:LiH-e1}
\end{figure}
\subsection{CEP effects in EI of $\mathrm{LiH}$}
In this section, we discuss the CEP effects in EI. The importance of the CEP of the few-cycle pulse 
in EI of polar molecules was demonstrated on a one-electron system, $\mathrm{HeH}{}^{2+}$ in 
Ref.~\cite{Kamta-PRL-05,Kamta-PRA-07} and on a two-electron system, $\mathrm{HeH}{}^{+}$ in 
Ref.~\cite{Dehghanian-JCP-13}. In this work we illustrate the importance of CEP effects in EI in
the four-electron $\mathrm{LiH}$ model. As alluded in the beginning of Sec.~\ref{numerical_ex},
to observe such CEP effects experimentally would require a pump-probe scheme.
The $\mathrm{LiH}$-like molecule can have two orientations w.r.t the linearly polarized laser field.      
\begin{figure}
 \includegraphics[width=\columnwidth]{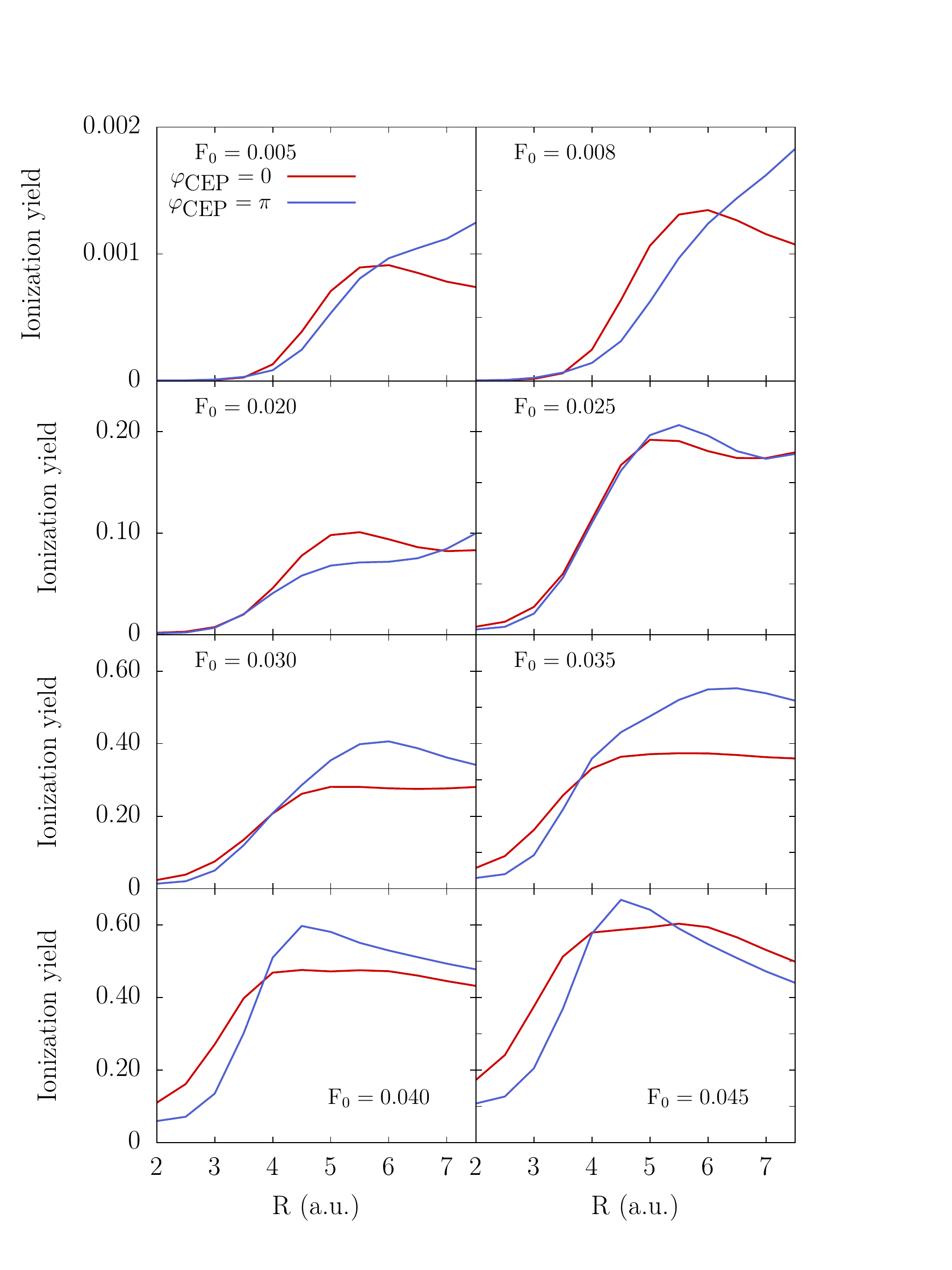}
 \caption{(color online). Ionization yield vs inter-nuclear distance $R$ for the $\mathrm{LiH}$ model 
          in the SAE approximation. See Fig.~\ref{fig:E_field}. for the definition of $\varphi_{\textup{CEP}}$.
          The panel with ${F}_0 = 0.005$ is multiplied by a factor of ten to scale. }
 \label{fig:LiH-sae}
\end{figure}
\begin{figure}
 \includegraphics[width=\columnwidth]{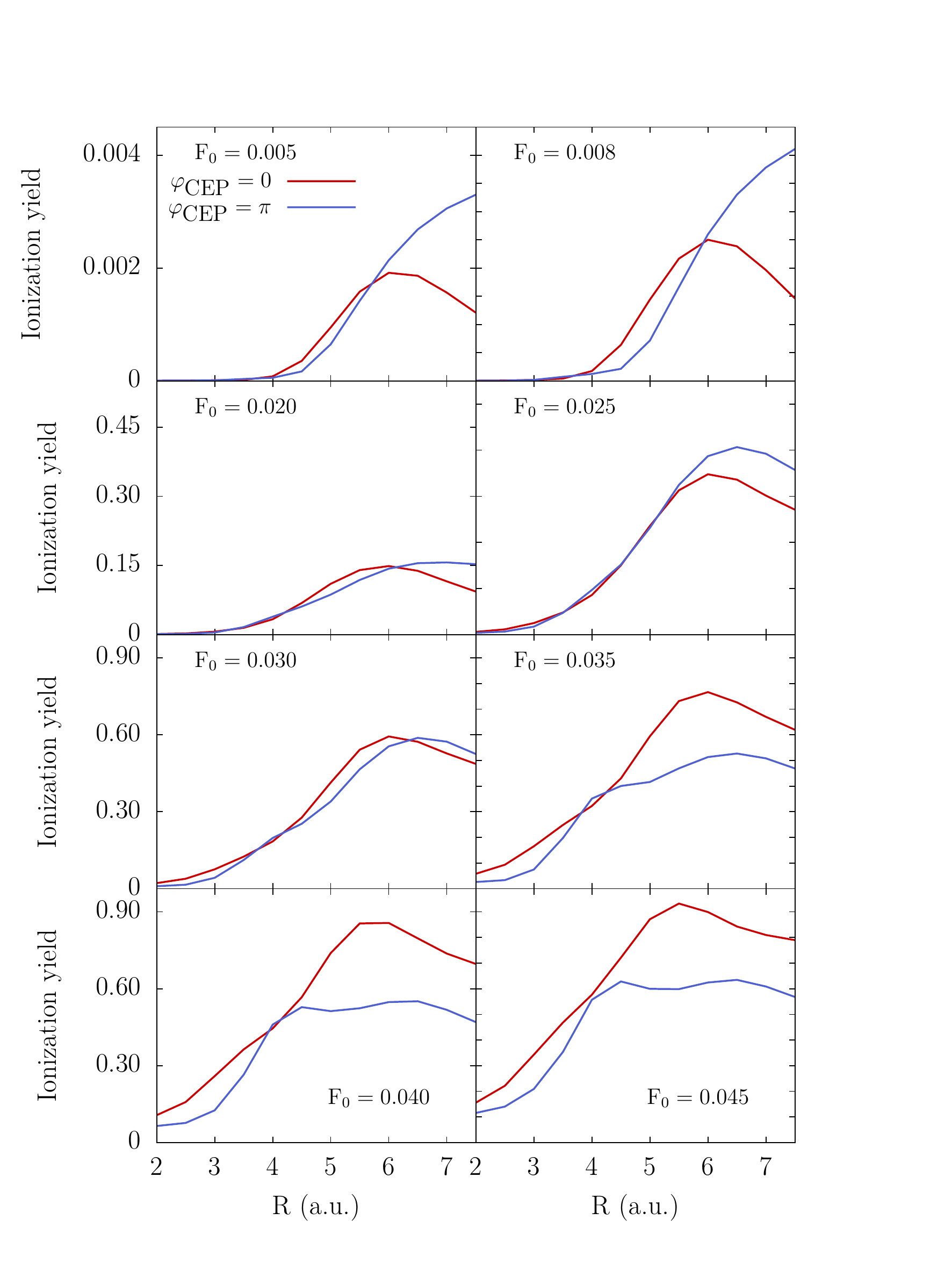}
 \caption{(color online). Ionization yield vs inter-nuclear distance $R$ for the $\mathrm{LiH}$ model 
          in the CAS$^*(2,5)$ model.  See Fig.~\ref{fig:E_field}. for the definition of $\varphi_{\textup{CEP}}$.
          The panel with ${F}_0 = 0.005$ is multiplied by a factor of ten to scale.}
  \label{fig:LiH-cas25}
\end{figure}
In Fig.~\ref{fig:LiH-e1}, we show a schematic diagram of $\mathrm{LiH}$ interacting with
a laser field. The interaction with ${F(t)}$ at the peak of the pulse with $\varphi_{\textup{CEP}} = 0$ 
[Fig.~\ref{fig:E_field}(a)] is illustrated in Fig.~\ref{fig:LiH-e1}(a) and with  
$\varphi_{\textup{CEP}} = \pi $ [Fig.~\ref{fig:E_field}(b)] in Fig.~\ref{fig:LiH-e1}(b), together 
with the field-free valence HF orbital. To study
the CEP effects, the intensities are chosen to illustrate the correlation effects in the EI 
phenomena in the tunneling as well as in the over-barrier-ionization regime. 
First we use ${F}_0 = 0.005$, $0.008$ which are below the over-barrier-field strength 
and represents the tunneling regime and the others with ${F}_0 = 0.020$, up to $0.045$, 
which are above the over-barrier-field strengths. 

The CEP effects in the EI are shown in Figs.~\ref{fig:LiH-sae} (SAE) and \ref{fig:LiH-cas25} [correlated CAS$^*(2,5)$].
In  the tunneling regime, the EI with $\varphi_{\textup{CEP}} = 0$ shows a completely different trend 
than with the $\varphi_{\textup{CEP}} = \pi$. The EI peak is well captured with 
$\varphi_{\textup{CEP}} = 0$, but with $\varphi_{\textup{CEP}} = \pi$ it is absent at the expected 
inter-nuclear separation. The same feature is observed with the CAS$^*(2,5)$ model in 
Fig.~\ref{fig:LiH-cas25}. The correlated model predicts a larger ionization yield at the critical 
inter-nuclear separation than the SAE approximation. It is important to emphasize that the EI peak 
is shifted towards larger inter-nuclear separation with $\varphi_{\textup{CEP}} = \pi$. This is 
a CEP effect because both, the SAE and the CAS$^*(2,5)$ approximations, predicts the same trend.
As we increase the field strength to  $F_0 = 0.025$, the SAE approximation predicts the EI 
peak for both $\varphi_{\textup{CEP}}$  at similar inter-nuclear separation. However, in the SAE 
approximation there is a cross-over between the EI peak with two different $\varphi_{\textup{CEP}}$.     
As we increase the intensity, the SAE approximation predicts more ionization yield with 
$\varphi_{\textup{CEP}} = \pi$ than with $\varphi_{\textup{CEP}} = 0$ at the critical inter-nuclear
separation. The correlated model gives a completely different picture. In this model, the 
ionization yield is larger for $\varphi_{\textup{CEP}} = 0$ than for  $\varphi_{\textup{CEP}} = \pi$
for $F_0 > 0.035$, which is certainly due to correlation effects.  The previous 
calculations with one- and two-electron molecules predict larger ionization yield with the 
$\varphi_{\textup{CEP}} = 0$ than with $\varphi_{\textup{CEP}} = \pi$ at the critical inter-nuclear 
separation~\cite{Kamta-PRL-05,Kamta-PRA-07,Dehghanian-JCP-13}. However, our studies show that there exists a 
cross-over between the $\varphi_{\textup{CEP}} = 0$ and $\varphi_{\textup{CEP}} = \pi$ cases in the
over-barrier regime as well as the $\varphi_{\textup{CEP}} = \pi$ predicts more ionization yield 
than with $\varphi_{\textup{CEP}} = 0$ in the tunneling regime. 
Therefore the CEP of a few-cycle pulse plays an important role in describing the direction of 
electron ejection in both tunneling and over-barrier-ionization, and most importantly the 
electron correlation can completely reverse the trend of the SAE approximation as seen from, e.g, 
the   $F_0 = 0.035$  case.

\section{Summary and Conclusion}
\label{conc}
In this work, we investigated the importance of electron correlation effects in enhanced ionization 
of model diatomic molecules by using the TD-GASCI method. The strength of the TD-GASCI method is to 
incorporate the electron correlation systematically. We used 1D models of $\mathrm{H}_2$ and 
$\mathrm{LiH}$ to study the electron correlation effects. We used the imaginary time propagation to 
calculate the ground state of the molecules. We then gave a detailed analysis of EI with different 
laser intensities and illustrated the features that arise because of electron correlation by using 
the TD-GASCI method. We found that correlation is very important to describe the EI behavior in the 
$\mathrm{H}_2$ model. The SAE approximation failed to describe  EI in $\mathrm{H}_2$ model and we 
would expect the same trends in real 3D systems. Another important conclusion from our study is 
the failure of the CIS approximation to correctly describe the EI phenomena. On the other hand the 
different GAS approximations predicted the correct behavior of EI and as we incorporated more and 
more correlation in the H$_2$ model the result converged towards the TDSE calculations. For 
the LiH model we found that EI 
persists and that the correlation shifts the peak of the EI towards 
larger  inter-nuclear separation. We also estimated the computational cost to solve the TD-GASCI 
equations with different GAS partitioning and found that it is computationally feasible and 
inexpensive compared to TDSE calculations. We emphasized the advantage of two active 
electrons over four active electrons in the GAS partition for the particular problem of
strong-field ionization where the valence electrons play a dominant role. Compared to the exact 
four electrons in the GAS partition, the two active electron model predicts the yield 
correctly and with much less computational cost. The CEP of a few-cycle pulse plays an important role 
in describing the EI in the asymmetric $\mathrm{LiH}$ model. The intriguing feature is the cross-over 
between the $\varphi_{\textup{CEP}} = 0$ and $\varphi_{\textup{CEP}} = \pi$ in the tunneling as well
as in over-barrier-ionization regime. In conclusion the electron correlation plays a crucial role in 
the EI phenomena.

\begin{acknowledgments}
This work was supported by the ERC-StG (Project No. 277767-TDMET), the VKR center of excellence, 
QUSCOPE, and the BMBF in the frame of the ``Verbundprojekt FSP 302''

\end{acknowledgments}
\bibliography{references}{}

\end{document}